\documentclass[twocolumn,prb]{revtex4}

\usepackage{graphicx}
\usepackage{rotating}
\usepackage{amsmath}
\usepackage{amsfonts}
\usepackage{amssymb}
\usepackage{enumerate}
\usepackage{longtable}
\usepackage{dcolumn}
\usepackage{bm}

\begin{document}

\newcommand{\be}{\begin{equation}}
\newcommand{\ee}{\end{equation}}
\newcommand{\bn}{\begin{eqnarray}}
\newcommand{\en}{\end{eqnarray}}

\title{$\alpha$-FeSe as an orbital-selective incoherent metal: 
An LDA+DMFT study}

\author{L. Craco,$^1$ M.S. Laad,$^2$ and S. Leoni$^1$}

\affiliation{$^1$Max-Planck-Institut f\"ur Chemische Physik fester 
Stoffe, 01187 Dresden, Germany \\
$^2$Technische Universit\"at Dortmund, Lehrstuhl f\"ur Theoretische 
Physik I, 44221 Dortmund, Germany}

\date{\rm\today}

\begin{abstract}
$\alpha$-FeSe, a prototype iron-chalcogenide superconductor, shows clear 
signatures of a strange incoherent normal state. Motivated thereby, we 
use LDA+DMFT to show how strong multi-band correlations generate a 
low-energy pseudogap in the normal state, giving an incoherent metal 
in very good semi-quantitative agreement with observations. We interpret 
our results in terms of $\alpha$-FeSe being close to {\it Mottness}.  
A wide range of anomalous responses in the ``normal'' state are 
consistently explained, lending strong support for this view. 
Implications for superconductivity arising from such an anomalous state
are touched upon.  
\end{abstract}
   

\maketitle

High Temperature Superconductivity (HTSC) in the recently discovered Iron
pnictides (FePn) is the latest surprise among a host of others in $d$- and
$f$ band materials.~\cite{[1]} While unconventional superconductivity
(U-SC) sets in close to the border of a frustration-induced~\cite{[2]} 
striped-spin-density-wave (SSDW) state with doping in the so-called 
$1111$-FePn, no magnetic long range order (LRO) is seen in the 
tetragonal $(\alpha)$ phase of Iron Selenide (FeSe)~\cite{[imai]} and 
FeSe$_{1-x}$Te$_{x}$,~\cite{fang} labelled $11$ systems, for small $x$
in ambient condictions.

The $11$- systems are structurally simpler than the $1111$- and the
$122$-FePn, without As or O. A rich variety of ground states reveal 
themselves upon external perturbations like doping, pressure and 
strain.~\cite{[str-fese]} Undoped $\alpha$-FeSe exhibits superconductivity
with $T_c=$9~K: upon applying pressure, $T_{c}$ dramatically rises to
$37$~K.~\cite{[imai]} U-SC is extremely sensitive to stoichiometry -
minute non-stoichiometry in Fe$_{1+y}$Se destroys SC.~\cite{[mcqueen]} 
U-SC at $T_{c}=34$~K is even observed in the high pressure 
{\it orthorhombic} structure in FeSe,~\cite{[nunez-reguiero]} in 
strong contrast to the $1111$-FePn, where it is only stable in the 
tetragonal structure. Interestingly, a two-step increase in $T_{c}$ 
as a function of pressure (with a large $dT_{c}/dP$ beyond $P_{c1}=1.5$~GPa), 
reminiscent of the $f$-electron U-SC, CeCu$_{2}$Si$_{2}$,~\cite{[steg]} is
observed.~\cite{[miyoshi]} In contrast, U-SC in FeSe is suppressed under 
tensile strain.~\cite{[brahimi]}  

In FeSe, the absence of charge
reservoir layers (in contrast to the 1111 and 122 FePn) leads to a
reduction in  $c$-axis length.  This has interesting consequences: in 
a correlated multi-band situation, changes in chemical composition are 
expected to sensitively affect the electronic and structural properties 
within Fe$_{2}$Se$_{2}$ layers,~\cite{will} changing the delicate balance 
between competing ordered states. This might explain the 
{\it extreme sensitivity} of the superconducting state to stoichiometry 
in FeSe.~\cite{[mcqueen]} Thus, one may ask, ``{\it how different}, or 
similar, is $\alpha$-FeSe from doped Iron arsenide 
superconductors?''~\cite{[imai]}  

Extant experiments for the normal state show clear strong correlation 
fingerprints. Photoemission (PES) experiments~\cite{PES,pes1} clearly 
evidence an incoherent, pseudogapped metallic state~\cite{PES} in
$\alpha$-FeSe, instead of a narrow Landau quasiparticle peak at $E_{F}$.  
Extant LDA calculations~\cite{pick} compare poorly with PES data, as is
checked by direct comparison (also see below).  In addition, the 
ultrahigh-resolution PES spectra show a low energy kink at 
$\approx 8~m$eV~(Ref.~\onlinecite{pes1}). As in 
1111-compounds,~\cite{SmLaFe,optc} this kink sharpens with cooling, 
and evolves smoothly across $T_{c}$. In contrast to the $1111$-FePn, in 
$\alpha$-Fe(Se$_{1-x}$Te$_{x})$, 
the antiferromagnetic (AF) ordering wave-vector, 
${\bf Q_{AF}}=(\delta\pi,\delta\pi)$,~\cite{bao} is very different from that 
predicted by LDA: this has an important consequence.  {\it If} SC is 
mediated by AF spin fluctuations,~\cite{[imai]} this implies that LDA is 
fundamentally inadequate to address magnetic fluctuations in the ``normal'' 
state. Depending upon $x$, SC either arises 
from an insulator-like normal state,or from a bad metal 
with $\rho_{dc}(T) \propto T$.~\cite{insul}  Further, NMR data~\cite{[imai]} 
show marked enhancement of antiferromagnetic (AF) spin fluctuations: no 
Korringa-like behavior is seen in $1/T_{1}$.  The uniform spin susceptibility 
anomalously increases for $T$ above $T_{c}$. The first two are reminiscent 
of those observed in high-$T_{c}$ cuprates up to optimal doping,~\cite{[pwa]} 
and the third is also found in the 1111-FePn as well as in another poorly 
understood U-superconductor system Na$_{x}$CoO$_{2}$.~\cite{[ong-nacoo2]} 
Finally, a minute amount of alloying by Cu drives $\alpha$-FeSe to a 
Mott-Anderson insulator.~\cite{[cava]}  Thus, $\alpha$-FeSe is close to a 
metal-insulator transition, i.e, to Mottness.  Needless to say, a proper 
microscopic understanding of the coupled charge-orbital-spin correlations 
manifesting in such anomalous behavior in $\alpha$-FeSe is a basic 
prerequisite for understanding how SC emerges from such a ``normal'' state. 
Extant theoretical understanding is restricted to one-electron band structure 
calculations.~\cite{pick}   

LDA based approaches are unable, by construction, to describe the incoherent 
metal features documented above. Here, we undertake a systematic LDA+DMFT 
study of $\alpha$-FeSe, and find that the electronic properties of this
layered superconductor are {\it partially} reminiscent of slightly underdoped 
1111-FePn superconductors. Sizable electronic correlations are shown to be
{\it necessary} for gaining proper insight into the anomalous normal state
responses in this system.  Very good semi-quantitative agreement with
PES (Ref.~\onlinecite{PES}) strongly supports this proposal.  Armed with 
this agreement, we analyze the non-Fermi-liquid (non-FL) metal in detail and 
predict specific anomalous features; these serve as a ``smoking gun'' for 
our proposal.  
 
\begin{figure}[thb]
\includegraphics[width=\columnwidth]{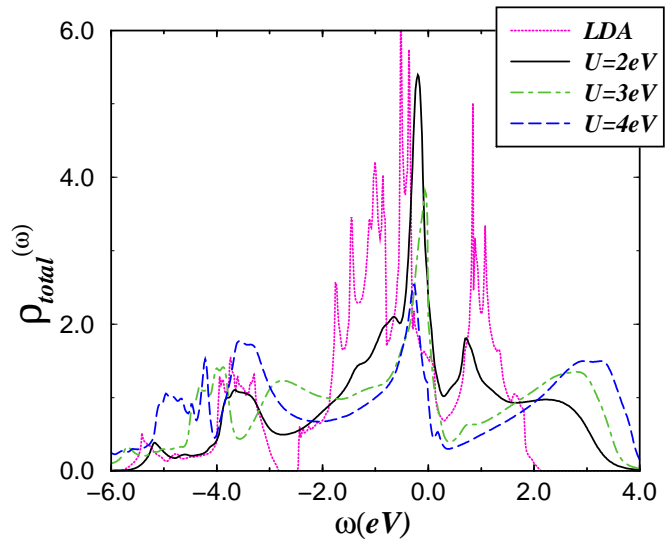}
\caption{
Comparison between the LDA (dotted) and LDA+DMFT (solid, dot-dashed
and long-dashed) density-of-states (DOS) for the Fe $d$ orbitals in 
$\alpha$-FeSe.  Large-scale transer of spectral weight from low energy 
to the Hubbard bands with increasing $U$ is visible. Also clear is the 
destruction of the low-energy Fermi liquid quasiparticle peak at $U=4$~eV.}
\label{fig1}
\end{figure}

We start with the tetragonal (space group: $P4/nmm)$ structure
of $\alpha$-FeSe with lattice parameters derived by Hsu {\it et al.} 
(Ref.~\onlinecite{hsu}). One-electron band structure calculations based
on local-density-approximation (LDA) were performed for $\alpha$-FeSe
using the linear muffin-tin orbitals (LMTO)~\cite{ok} scheme. 
Our LDA results for the total density of states (DOS) is shown in 
Fig.~\ref{fig1} (dotted line). Similar total DOS 
were also obtained by other groups,~\cite{pick} showing that the 
electronic states relevant to Fe-superconductors are Fe $d$-band states. 
As found in previous calculations, the Fe-$d$ bands hybridize with 
Se-$p$ bands around -3.8~eV, giving rise to a small, separated band 
below 3~eV binding energy. Interestingly, the resulting ``gap'' at high 
energy is not seen in PES experiments,~\cite{PES,pes1} which show 
only a broad continuum in this energy range. As discussed below, this
discrepancy is resolved by dynamical spectral weight transfer (SWT) 
which originates from sizable electronic correlations in FeSe.  

Though LDA provides reliable structural information on a one-electron 
level, it generically fails to capture the ubiquitous dynamical 
correlations in $d$-band compounds, and so {\it cannot} access normal
state incoherence in $d$-band systems. Combining LDA with 
dynamical-mean-field-theory (DMFT) is the state-of-the-art prescription 
for remedying this deficiency.~\cite{[kotliar]} Within LDA, the one-electron
part for $\alpha$-FeSe is
$H_{0}=\sum_{{\bf k},a,\sigma}
\epsilon_{a}({\bf k})c_{{\bf k},a,\sigma}^{\dag}c_{{\bf k},a,\sigma}\;,$
where $a=x^{2}-y^{2},3z^{2}-r^2,xz,yz,xy$ label the diagonalized, five
$d$ bands. Further, in light of the strong correlation signatures cited 
above, the full, multi-orbital (MO) Coulomb interactions must be included. 
These constitute the interaction term, which reads
$H_{int}=U\sum_{i,a}n_{ia\uparrow}n_{ia\downarrow}
+ U'\sum_{i,a \ne b}n_{ia}n_{ib}
-J_{H}\sum_{i,a,b}{\bf S}_{ia}.{\bf S}_{ib}$.
To pinpoint the relevance of sizable MO electronic interactions in the 
system, we present LDA+DMFT results for $U=2,3,4$~eV, $U'=U-2J_H$~eV, 
and fixed $J_{H}=0.7$~eV.  To solve the MO-DMFT equations, we use the MO 
iterated-perturbation-theory (IPT) as an impurity solver.~\cite{SmLaFe,ePAM} 
Though not quantitatively exact, this solver is numerically very efficient, 
is valid at $T=0$, and self-energies $[\Sigma_a(\omega)]$ can be computed 
very easily.  Given the complexity in FeSe with five $d$ bands, these are 
particularly attractive features not shared by more exact solvers.

We now present our results for the normal phase of $\alpha$-FeSe.  
Fig.~\ref{fig1} shows how LDA+DMFT modifies the LDA band structure.  
MO dynamical correlations arising from $U,U'$ and $J_{H}$ lead
to spectral weight redistribution over large energy scales and
the formation of lower- (LHB) and uper-Hubbard (UHB) bands. As
seen, the UHB at 2.4~eV for $U=2$~eV (and, $U'=0.6$~eV) moves to
higher energies with increasing $U$.  The LHB is not clearly resolved 
$U \leq 2$~eV. Indeed, we observe a relatively sharp and quasi-coherent 
low-energy peak, with a prominent shoulder feature instead of the LHB 
at $\omega_{L}\simeq -1.0$~eV.  Similar features are visible in other  
results (Ref.~\onlinecite{anisi}) for similar $U$ values. Correlation 
effects, however, become more visible at $U \geq 3$~eV. In contrast to 
the $U=2$~eV result, a LHB at 2.8~eV binding energy is clearly resolved 
with $U=3$~eV. With increasing $U$, the LHB is shifted toward energies 
where the Se-$p$ bands occur in the LDA. Interestingly, this superposition 
of the $pd$-band and LHB for $U=4$~eV makes it impossible to cleanly 
resolve the LHB in PES. Hence, estimation of the degree of correlatedness 
in FeSe cannot be based solely on the ``absence'' of the LHB in PES, and 
must involve deeper analysis of PES, in conjuction with other probes, 
before a definitive conclusion can be drawn.   

Fig.~\ref{fig1} shows that the DOS at $E_F$ is pinned to its LDA value for 
$U \leq 3$~eV.  This is the expected behavior for a FL metal.~\cite{anisi}  
With increasing $U$, however, our LDA+DMFT results show drastic 
modification of the spectral functions near $E_F$. Revealingly, in 
addition to large-scale SWT, we find that the FL-like pinning of the 
LDA+DMFT DOS to its LDA value, found for small $U$, is lost for
$U=4$~eV. Instead the metallic state shows a clear pseudogap at $E_{F}$, 
with no Landau FL quasiparticles (see below). A related bad metallic state 
has also been found in earlier LDA+DMFT works~\cite{SmLaFe,optc} for the 
1111-FePn, and, as shown there, is in very good semi-quantitative 
agreement with a host of experimental observations.

 \begin{figure}[thb]
\includegraphics[width=\columnwidth]{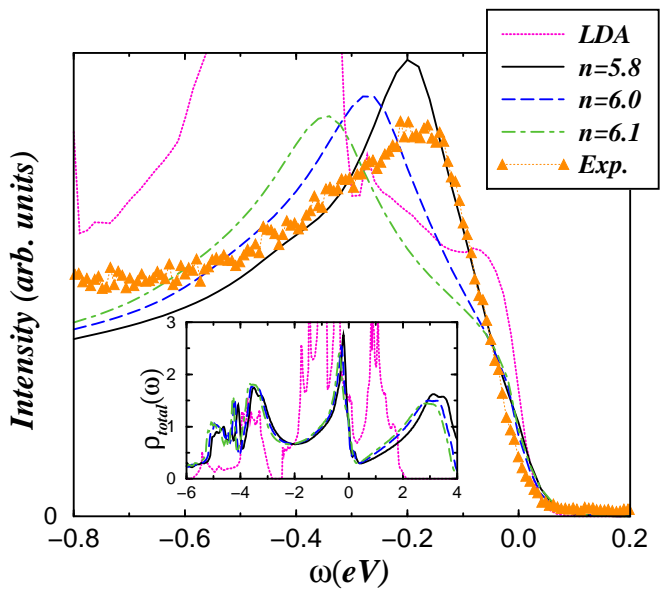}
\caption{
Comparison between the LDA+DMFT result for FeSe and angle-integrated 
photoemission (PES, triangles).~\cite{PES} Very good quantitative 
agreement is clearly seen for $n=5.8$. In particular, the low-energy 
energy spectrum (up to 0.1~eV binding energy) {\it and} the peak at 
$-0.19$~eV in PES is accurately resolved in the DMFT spectrum with 
$U=4.0$~eV.  Clearly, LDA spectrum compare poorly with PES.
(The inset shows the total LDA+DMFT spectral functions. LDA result is
shown for comparison.)}
\label{fig2}
\end{figure}

In Fig.~\ref{fig2}, we compare our $U=4$~eV (and, $U'=2.6$~eV) results with
PES for doped FeSe$_{1-x}$.~\cite{PES} Very good semi-quantitative
agreement with experiment is visible for $n=5.8$, where $n$ is the total
band filling of the $d$ shell. In particular, the broad peak at
$\approx -0.17$~eV in PES is faithfully reproduced by LDA+DMFT.
(Comparison with Fig.~\ref{fig1} also shows clear disagreement between 
PES and the LDA as well as $U \leq 3$~eV results.) For comparison, the 
computed LDA+DMFT spectra for the undoped ($n=6.0$) and electron doped 
($n=6.1)$ cases show clear disagreement with PES at low energies. In 
contrast to this, the correlated spectral functions close to $E_F$ are 
insensitive to (small) changes in the electron (hole) concentration; we 
predict that combined PES/XAS on doped-FeSe samples will show this in 
future.  Interestingly, we see that, in contrast to the PES spectra, 
XAS lineshapes are less sensitive to $3.0 \leq U\leq 4.0$~eV (see 
Fig.~\ref{fig1}). Recall that we obtain a correlated FL for $U=3.0$~eV 
going over to an incoherent metal for $U=4.0$~eV.  We suggest, therefore, 
that inspection of XAS spectra alone~\cite{[brink]} is inadequate to 
address the issue of the degree of correlations in the Fe pnictides 
in general, and, to do so, one must consider the {\it full} one-particle 
spectral function via PES+XAS data taken together.

\begin{figure}[thb]
\includegraphics[width=\columnwidth]{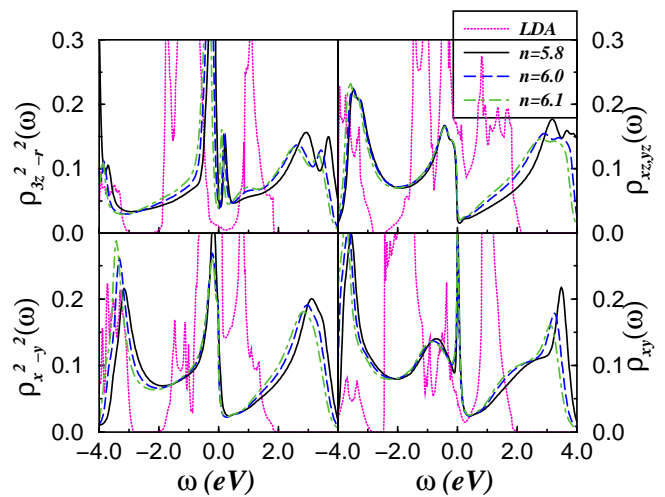}
\caption{
Orbital-resolved LDA (dotted) and LDA+DMFT (with $U=4.0$~eV, $U'=2.6$~eV and
$J_{H}=0.7$~eV) density-of-states (DOS) for the Fe $d$ orbitals in FeSe for 
three doping values. Large-scale dynamical spectral weight transfer occuring 
hand-in-hand with orbital selective incoherence is clearly visible.}
\label{fig3}
\end{figure}

\begin{figure}[thb]
\includegraphics[width=\columnwidth]{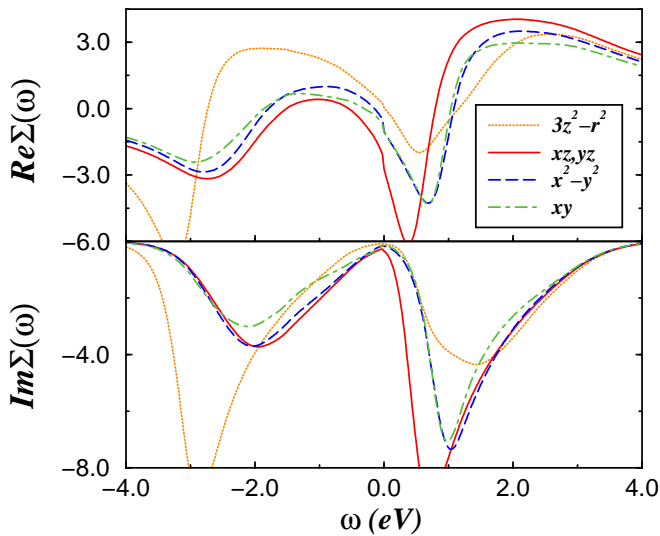}
\caption{
Orbital-resolved LDA+DMFT self-energies for electron-doped $\alpha$-FeSe. 
Upper panel: Real parts, clearly showing a low-energy kink feature, about 
$15~m$eV below $E_{F}$, in Re$\Sigma_{a}(\omega)$ with $a=xy,xz,yz,x^{2}-y^{2}$.
Lower panel: The corresponding imaginary parts, showing clear
sub-linear ($xy,xz,yz,x^{2}-y^{2}$) and almost quadratic ($3z^{2}-r^{2}$)
in-$\omega$ dependence for $\omega \leq E_{F}$. }
\label{fig4}
\end{figure}

We now focus on orbital resolved spectral functions of $\alpha$-FeSe. Clear 
orbital-selective (OS) incoherence is visible in Fig.~\ref{fig3}: a 
low-energy pseudogap is clearly  visible in the $xz,yz,x^{2}-y^{2}$ DOS, 
and only the $xy,3z^{2}-r^{2}$ DOS show very narrow FL-like resonances at 
$E_{F}$. Examination of the self-energies in Fig.~\ref{fig4} shows that, 
for $n=5.8$, only Im$\Sigma_{3z^2-r^2}(\omega)\simeq -a\omega^{2}$ for 
$\omega <E_F(=0)$. Using the Kramers-Kr\"onig relation, it follows that 
the Landau FL quasiparticle residue, $Z$ vanishes near-identically for the 
$xz,yz,x^{2}-y^{2}$ band carriers [from Re$\Sigma (E_F)$], direct numerical 
evaluation gives $Z_{xz,yz}=0.023, Z_{x^{2}-y^{2}}=0.04$). Correspondingly, 
spectral lineshapes for these bands are nicely fit by a power-law fall-offs 
(not shown) in the range $-2.0<\omega<-0.2$~eV; this local ``critical'' 
behavior is cut-off by the normal state pseudogap for 
$-0.2<\omega < E_{F}$. Hence, at small but finite $T$, the ``normal'' 
metal will be totally incoherent, without any FL quasiparticles. Remarkably, 
such behavior results from strong scattering between {\it effectively} 
(Mott) localized and itinerant components of the full DMFT matrix 
propagators, and is caused by an Anderson {\it orthogonality} catastrophe 
(AOC) in the impurity problem of DMFT. This is intimately linked to OS 
Mott-like physics within DMFT.~\cite{[georges]}

Our identification of normal state incoherence in $\alpha$-FeSe with the 
AOC has many interesting consequences.  Since the optical conductivity
($\sigma(\omega)$) in DMFT is a direct convolution of the full one-particle 
propagators, we predict that $\sigma(\omega)$ should show a pseudogapped form
at small $\omega$, followed by a smooth crossover to a power-law 
($\simeq\omega^{-\eta}$) behavior at higher energy.  The $dc$ resistivity at 
``high'' $T$ will be controlled by the renormalized scattering rate,
$\tau^{*}(\omega)^{-1}=\omega \frac{{\rm Re} \sigma(\omega)}{\rm Im \sigma(\omega)} \simeq \omega$.~\cite{[vdm-nature]}  Thus, $\rho_{dc}(T)\simeq T$ at ``high'' $T$, 
as is ubiquitous to FeSe for $T>T_{c}$. Using the 
Shastry-Shraiman~\cite{[shastry]} relation relating the $B_{1g}$ electronic 
Raman scattering (ERS) intensity to the optical conductivity, we predict 
that the ERS lineshape will also show a low-energy pseudogap, followed
by a weakly $\omega$-dependent continuum at higher energy.  These are
stringent tests for our proposal, and experimental verification would 
place it on solid ground.

Also, the extreme sensitivity to Cu doping, which drives FeSe to a Mott
insulator,~\cite{[cava]} is readily rationalized in our picture.  In an
incoherent metal with singular or near singular behavior of the one-particle
propagator, disorder is a strongly {\it relevant} perturbation, and minute
concentration of impurities qualitatively changes the low-$T$ behavior of the
system from an incoherent metal to a kind of Anderson-Mott 
insulator.~\cite{[varma]}  We emphasize that such sensitivity to minute 
impurity concentration is neither expected, nor found, in a weakly correlated 
FL.  As it turns out, this is also additional evidence for $\alpha$-FeSe 
being close to Mottness.

Finally, our finding of an incoherent non-FL state implies that interband
one-electron mixing is {\it irrelevant} in the normal state, since single
{\it electrons} cannot {\it coherently} tunnel between different orbitals 
in such a metal.  In analogy with coupled Luttinger liquids,~\cite{[pwa]} 
{\it two-particle} coherence (arising from a second-order process involving 
interband one-particle mixing) should then take over.  As $T$ is lowered, 
therefore, various two-particle instabilities, either in the particle-hole 
(magnetism) or particle-particle (superconductivity) sector, will 
destabilize such a non-FL metal.  Detailed consideration of these 
instabilities and such a mechanism for SC is out of scope of this
work,~\cite{[our-PRL]} and is left for the future.  

To conclude, based on a first-principles LDA+DMFT study, we have shown 
that orbital-selective incoherence characterizes the ``normal'' metallic 
phase in $\alpha$-FeSe.  Very good semi-quantitative agreement with PES 
spectra and rationalization of a variety of unusual observations in a 
single picture lend strong support for our proposal.  Sizable multi-orbital 
correlations are shown to be {\it necessary} to derive this orbital-selective 
incoherent metal. Emergence of SC at low $T$, along with extreme sensitivity 
of the ground state(s) to minute perturbations in FeSe$_{1-x}$Te$_{x}$ or 
Cu$_{y}$Fe$_{1-y}$Se should thus be considered as some manifestations of 
the myriad possible instabilities of such an incoherent non-Fermi liquid 
metal in close proximity to a Mott insulator. 

S.L. acknowledges ZIH Dresden for computational time.

\end{document}